\documentclass[11pt,aps,prc,superscriptaddress,showpacs,floatfix]{revtex4}
\usepackage[dvips]{graphicx}
\usepackage{epsfig}
\usepackage{bm}
\usepackage{amssymb}
\usepackage{wrapfig}
\usepackage{ulem}
\usepackage{multirow}
\usepackage{color}

\newcommand{\epsi}{\mbox{$\varepsilon$}}

\newcommand{\vp}{\mbox{$\bm{p}$}}

\newcommand{\vbr}{\mbox{$\bm{r}$}}
\newcommand{\vk}{\mbox{$\bm{k}$}}
\newcommand{\vn}{\mbox{$\bm{n}$}}

\newcommand{\vB}{\mbox{$\bm{B}$}}

\newcommand{\vR}{\mbox{$\bm{R}$}}

\begin{document}

\title{\bf Rapid Spin Deceleration of Magnetized Proto-Neutron
Stars \\
via Asymmetric Neutrino Emission}

\author{Tomoyuki~Maruyama}
\affiliation{College of Bioresource Sciences,
Nihon University,
Fujisawa 252-8510, Japan}
\affiliation{National Astronomical Observatory of Japan, 2-21-1 Osawa, Mitaka, Tokyo 181-8588, Japan}

\author{Jun~Hidaka}
\affiliation{National Astronomical Observatory of Japan, 2-21-1 Osawa, Mitaka, Tokyo 181-8588, Japan}

\author{Toshitaka~Kajino}
\affiliation{National Astronomical Observatory of Japan, 2-21-1 Osawa, Mitaka, Tokyo 181-8588, Japan}
\affiliation{Department of Astronomy, Graduate School of Science, University of Tokyo, Hongo 7-3-1, Bunkyo-ku, Tokyo 113-0033, Japan}

\author{Nobutoshi~Yasutake}
\affiliation{ Department of Physics, Chiba Institute of Technology, 
2-1-1 Shibazono, Narashino, Chiba 275-0023, Japan}

\author{Takami~Kuroda}
\affiliation{National Astronomical Observatory of Japan, 2-21-1 Osawa, Mitaka, Tokyo 181-8588, Japan}

\author{Tomoya~Takiwaki}
\affiliation{National Astronomical Observatory of Japan, 2-21-1 Osawa, Mitaka, Tokyo 181-8588, Japan}

\author{Myung-Ki Cheoun}
\affiliation{National Astronomical Observatory of Japan, 2-21-1 Osawa, Mitaka, Tokyo 181-8588, Japan}
\affiliation{Department of Physics, Soongsil University, Seoul, 156-743, Korea}

\author{Chung-Yeol Ryu}
\affiliation{General Education Curriculum Center, Hanyang University, Seoul, 133-791, Korea}

\author{Grant J. Mathews}
\affiliation{Center of Astrophysics, Department of Physics,
University of Notre Dame, Notre Dame, IN 46556, USA}

\date{\today}

\pacs{25.30.Pt, 26.60.-c, 24.10.Jv, 97.60.Jd}

\begin{abstract}
We estimate the maximum possible contribution to  the early spin deceleration of  proto-neutron stars due  asymmetric neutrino absorption.  
We calculate the neutrino scattering  in the context of a fully relativistic mean
field theory and estimate for the first time the spin deceleration of neutron stars due to asymmetric neutrino absorption in a toroidal magnetic field configuration. 
We find that the deceleration can be  much larger 
for asymmetric neutrino absorption in a toroidal magnetic field than
the  braking due to  magnetic dipole radiation.  Nevertheless, the effect is estimated to be  less than the angular momentum loss due to the transport of magnetically locked material in the neutrino energized wind. 
\end{abstract}

\maketitle
%\tableofcontents

\section{Introduction}

Magnetic fields play an important role in many astrophysical phenomena.
The observed  asymmetry in supernova (SN) remnants, pulsar kick
velocities~\cite{lyne94}, and the existence of
magnetars~\cite{pac92,mag3}  all suggest  that  
strong magnetic fields affect the dynamics of core-collapse  SN explosions 
and the  velocity~\cite{rothchild94}
that  proto-neutron stars (PNSs) receive at  birth.

There are at least two major SN explosion scenarios leading to asymmetric  morphologies 
in  observed SN remnants.
One of them is the standing accretion shock instability (SASI)-aided neutrino 
driven explosion~\cite{marek09,takiwaki12}.  
The other is the magneto-rotational explosion (MRE)~\cite{burrows07,kotake04}.
Both mechanisms may  also be a  source 
for  pulsar kick velocities~\cite{sawai08,nordhaus10}.
The MRE takes place through the extraction of the rotational energy 
of the PNS via strongly amplified magnetic fields $\sim10^{15}$G.
In this case a few processes can be candidates for the amplification 
mechanisms such as the winding effect or the magneto-rotational 
instability \cite{akiyama03}.
Thus, the MRE is expected to leave behind a  magnetar remnant.
However, there are many unknown aspects of these scenarios, 
such as the  progenitor and final core  rotation and magnetic field profile.  
Hence, there is not yet a definitive understanding of the observed asymmetry 
and remnant kick velocities in core collapse supernovae.

Moreover, it has been pointed out   \cite{nakano11} that the characteristic spin down ages ($P/2 \dot P$) of magnetars 
appear to be  systematically overestimated compared to
ages of the associated supernova remnants.  This suggests that there may be  
additional loss of angular momentum, perhaps  due \cite{nakano11}  to the 
dissipation of rotational energy into magnetic dipole radiation.  
It has also been proposed \cite{TCQ04} that magnetic proto-neutron stars can lose angular momentum from
the ejection of magnetically coupled material via  neutrinos.  However, there are alternative explanations  
for the spin down of magnetic neutron stars as summarized  in Refs.~\cite{Heger05,Ott06}.
Even in the non-magnetized case a rotating neutron star will lose 
considerable amounts of angular momentum by neutrino emission 
as first pointed out in Refs.~\cite{Kazanas77, Epstein78} and discussed 
in more detail in \cite{Baumgarte98}, and applied to a discussion of 
neutron star birth properties by Refs.~\cite{Janka04, Heger05,Ott06}.
 
Nevertheless, in this  work we point out that there is  yet another
source of angular momentum loss via neutrino  emission in magnetic
stars.  
In this case it is due to  asymmetric neutrino scattering 
in strong toroidal magnetic fields.
In this work we estimate the maximum possible effect from this
asymmetric scattering and compare it specifically with the spin down  
calculated by the mechanism of \cite{TCQ04}.  
We find that even in the best case, this contribution to spin down  is
less than that of other mechanisms.  
Nevertheless, it does contribute as an independent possible process and
one should consider this effect in models for the early spin down of 
proto-neutron stars.

Although we will approach obtaining estimates in a simple best case model,
one should keep in mind that this effect should be studied in the context of more complex neutron star models (e.g. \cite{Lat04,Lat12} and Refs. therein).
Both static and dynamic properties of  neutron-star matter have been studied 
(e.g.~Refs.~\cite{bb94,pb97,g01}) at high temperature and density in the context of  spherical non-magnetic neutron star models.  
 Such aspects as an
exotic phase of strangeness condensation (e.g.~\cite{kn86,K-con,bkpp88}),
nucleon superfluidity~(cf.~\cite{pb90}), rotation-powered thermal 
evolution (e.g.~\cite{t98}), a quark-hadron phase transition~(e.g.~\cite{ny09}), etc.,
have been considered. 
Neutrino propagation has also been studied for PNS matter including hyperons
(cf.\cite{rml98}).
These theoretical treatments of high-density
hadronic matter, however,  have not yet considered the effects of strong magnetic fields.

Although previous work (e.g. Refs.~\cite{arras99,lai98}) has studied
the effects of magnetic fields on the asymmetry of neutrino emission,
the neutrino-nucleon scattering processes were calculated
in a non-relativistic  framework \cite{arras99} and only a uniform dipole field configuration was considered.

Our studies of neutrino scattering and absorption cross sections 
in hot and dense magnetized neutron-star matter (including hyperons)~\cite{MKYCR11,MYCHKMR13}
are based upon the fully relativistic mean field (RMF) theory~\cite{serot97}.
Our previous papers demonstrated that poloidal magnetic fields enhance the  scattering cross-sections for  neutrinos
in the direction parallel to the magnetic field, 
while also reducing the absorption cross-sections in the same direction.
When the direction is anti-parallel, the opposite occurs.

It was shown in Ref.~\cite{MYCHKMR13} that forinterior magnetic field 
strengths near the equipartition limit, where by equipartition we
mean that magnetic pressure $\approx$ gas pressure.
This occurs for field strengths of order  $10^{16-18}$G. 
For such field strengths 
the enhancement of the scattering cross-sections is  
$\sim 1\%$ at a baryon density of $\rho_B= 3\rho_0$,
while the reduction in  the absorption cross section is $\sim 2\%$.
This  enhancement and reduction were shown to increase the neutrino momentum flux 
emitted along the direction of the dipole magnetic field and to decrease the emitted
momentum flux emitted antiparallel to the magnetic field. 
This asymmetry was then applied to a calculation of pulsar-kick velocities 
in the context of a one-dimensional  Boltzmann equation 
including only the dominant effect of neutrino absorption.
PNS kick velocities of $\sim 550$ km~s$^{-1}$ were estimated.

Of relevance to the present work, however, are recent magneto-hydrodynamic (MHD) PNS simulations 
(e.g.~\cite{BrSp04,TKS09,KM10}) which demonstrate   
that the magnetic field inside a neutron star can obtain a  toroidal configuration.
It was also demonstrated~\cite{TKS09} that the field strength
of toroidal magnetic field is $\sim$100 times stronger
than that of a poloidal magnetic field due to winding effects on the 
original dipole field lines for rapidly rotating  the proto-neutron stars.

Here, we show that the early  spin deceleration of a  PNS could result 
from an asymmetry in the   neutrino emission
that arises  from parity violation 
in weak interactions \cite{vilenkin95,horowitz98}
and/or an asymmetric distribution of the magnetic field \cite{bisnovat93}
in strongly magnetized PNSs.
(However, for an alternative scenario see Ref.~\cite{TCQ04}.)
Theoretical calculations \cite{arras99,lai98} have suggested 
that as little  $\sim$1\% asymmetry in the neutrino emission out of 
a total neutrino luminosity of $\sim 10^{53}$ ergs is  enough
to explain the observed pulsar kick velocities.  
We here study the asymmetric neutrino absorption 
in the case of  a toroidal magnetic field inside a proto-neutron star. 
If  neutrinos are preferentially  emitted
along a  direction opposite to that of the rotation.    
This could enhance the spin down rate of  PNSs.
In this article, we present for the first time  a study of the
effect of  asymmetric neutrino absorption on the spin deceleration of PNSs.

\section{The Model}

\subsection{Neutrino Transport in Relativistic Mean-Field Theory}

Even a strong magnetic field has less mass-energy  than the baryonic 
chemical potential in degenerate neutron-star matter, 
i.e.~$\sqrt{e B} \ll \epsi_b - M_b$, where $\epsi_b$ and $M_b$ are the chemical potential and rest mass of the baryon $b$, respectively.
Hence,  we can treat the magnetic field as a perturbation.
We then ignore  the contribution from  convection currents  and
consider only the spin-interaction.
We also assume that  $|\mu_b B| \ll E_b^*(\vp) = \sqrt{\vp^2 + M_b^{*2}}$,
and treat the single particle energies and the wave function
in a perturbative way.

In this framework we then obtain the wave function in a magnetic field
by solving the Dirac equation:
\begin{equation}
\left[ \gamma_\mu p^\mu - M^*_b -  U_0 (b) \gamma_0 - \mu_b B \sigma_z \right ] u_b(p,s) =0 ,
\label{DiracE}
\end{equation}
where $M^*_b = M_b - U_s (b)$,
while $U_s(b)$ and $U_0(b)$ are respectively the scalar mean-field and
the time-component of the vector mean-field for the baryons $b$.
These scalar and vector fields are calculated in the context of RMF theory.

In Refs.~\cite{MKYCR11,MYCHKMR13}
 we calculated the neutrino absorption cross-section $\sigma_A$
in PNS matter for an interior  magnetic field strength near equipartition
and a temperature of  $T = 20$ MeV.
Those results  demonstrated  that the absorption cross-sections are suppressed 
in the direction parallel to the magnetic field ${\bm B}$
by  about $2 - 4$\% in the density region of $\rho_B = (1 - 3) \rho_0$.
The opposite effect occurs  in  the anti-parallel direction.  
The net effect of these changes in the absorption cross sections leads to
an increase in  the emitted neutrino momentum flux along the direction 
of the magnetic field and a decrease of the momentum flux emitted in the antiparallel direction.

However,  it is quite likely  \cite{BrSp04,TKS09,KM10} that the magnetic field exhibits 
a toroidal configuration within the PNS.
Hence, we now  consider the implications of a toroidal field configuration
on  neutrino transport in a strongly magnetized PNS.
For this purpose we solve for  the neutrino phase-space distribution function
 $f_\nu (\vbr,\vk)$ using  a Boltzmann
equation as described below and in Ref.~\cite{MYCHKMR13}.

We assume that the system is static and nearly in local thermodynamic equilibrium.
Under these assumptions the phase-space distribution function satisfies 
$\partial f_\nu /\partial t = 0$ and can be expanded  as
$ f_\nu (\vbr,\vk) = f_0 (\vbr,\vk) +  \Delta f (\vbr,\vk)$, 
where the first term is the local equilibrium part, and the second term is
its deviation.

Furthermore, we assume that only the dominant effect of absorption contributes 
to the neutrino transport, 
and that the neutrinos travel along a straight line.  
It is common (e.g. \cite{burrows07}) to utilize a one-dimensional 
Boltzmann equation in simulations of PNS formation, and hence, the straight line approximation is adequate for our purpose.
The 1D Boltzmann equation for $f_{\nu}$  
in our simulation can then be written:
%\begin{equation}
\begin{eqnarray}
{\hat k} \cdot \frac{\partial}{\partial \vbr} f_\nu(\vbr,\vk)
&=& {\hat k} \cdot \frac{\partial \epsi_\nu (\vbr)}{\partial \vbr}
\frac{\partial f_0}{\partial \epsi_\nu}
+ {\hat k} \cdot \frac{\partial \Delta f}{\partial \vbr}
%\nonumber \\ &\approx&
\approx  - \frac{\sigma_{A}(\vbr,\vk)}{V} \Delta f (\vbr,\vk) ,
\label{Boltz1}
\end{eqnarray}
%\end{equation}
where $\epsi_\nu(\vbr)$ is
the neutrino chemical potential at  coordinate $\vbr$.
Here,  we define the variables
$x_L \equiv (\vbr \cdot \vk)/|\vk|$ and
$\vR_T \equiv \vbr - (\vbr \cdot \vk)\vk/\vk^2$, 
and then solve Eq.~(\ref{Boltz1}) analytically 
\begin{equation}
\Delta f(x_L,R_T,\vk) = \int_0^{x_L} d y
\left[ - \frac{\partial \epsi_\nu}{\partial y}
\frac{\partial f_0}{\partial \epsi_\nu} \right]
\exp\left[ - \int^{x_L}_{y} d z
\frac{\sigma_{A}(z, R_T, \vk)}{V} \right] ,
\label{BolSol}
\end{equation}
where  the center of the neutron-star
is at $\vbr = (0,0,0)$, and
all of the integrations are performed along
a straight line.

This simplified Boltzmann equation is adequate for our purpose which is to estimate the relative difference between scattering aligned with the magnetic field vs.~anti-aligned.  When the neutron star is rotating, however,
the neutrino transport  should to be treated in the comoving
frame of the fluid. This causes additional angular momentum loss as we discuss below.

In this work we utilize an equation of state (EOS) at a fixed temperature and 
lepton fraction by 
using the parameter set PM1-L1 \cite{kpcon} for the RMF as 
in  previous work \cite{MKYCR11,MYCHKMR13}.
When Lambda particles are  not included, 
the PM1-L1 EOS is sufficiently stiff \cite{K-con} to give a maximum  
neutron star mass with about 2.2 solar-mass 
which is  larger than the value observed for PSR J1614-2230 \cite{T-Solar}.
When the Lambda particles are  included, however, the EOS becomes softer and 
gives about 1.7 solar-mass as a maximum neutron-star mass.
This could be resolved, however,  if  we were to introduce additional repulsive force 
between the $\Lambda$s \cite{Bednarek11} consistent with hypernuclear data.  
Another possibility would be introducing a repulsive three-body nuclear force.

\begin{wrapfigure}{r}{8.5cm}
\begin{center}
\vspace{-0.5cm}
\hspace{0.5cm}
{\includegraphics[scale=0.45]{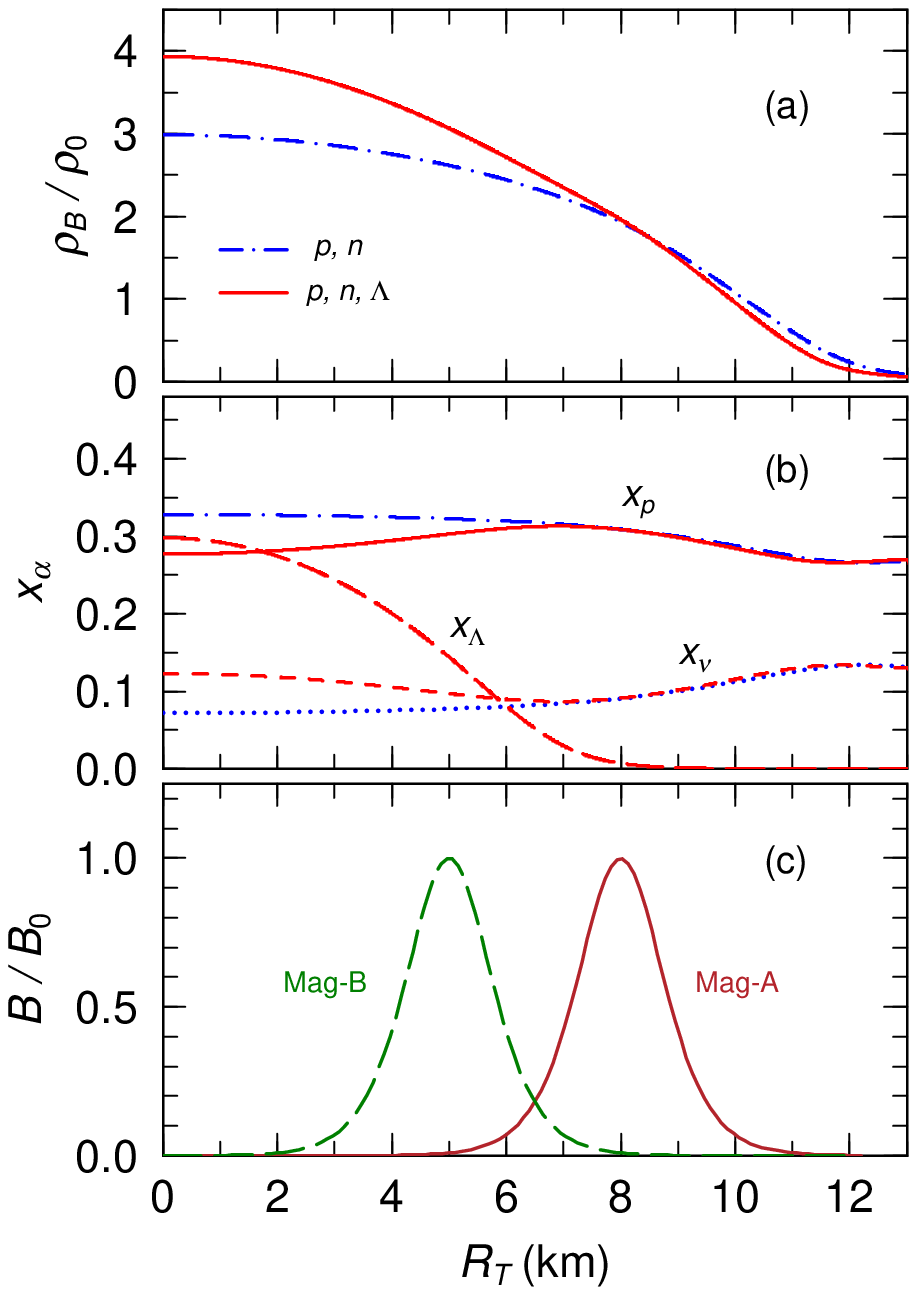}}
\caption{\small (Color online) 
The upper panel (a) shows the baryon density distribution
for a  PNS with $T$ = 20 MeV and $Y_L = 0.4$.
The solid and long-dashed lines show 
 results  with (red solid line) and without (blue dot-dashed line)  
$\Lambda$s, respectively.
The middle panel (b) shows the number fractions for protons, Lambdas and 
neutrinos in a PNS.
The (red) solid and (blue) dot-dashed  lines show the proton fraction 
in systems with and without $\Lambda$s.
The (red) long-dashed line indicates the Lambda fraction.
The (red) dashed and (blue) dotted lines denote the neutrino fraction 
in systems with  and without  $\Lambda$s, respectively.
The lower panel (c) shows the field strength distribution at $z=0$ 
for the toroidal magnetic fields considered here. 
The (wine red) solid and (dark green) dashed lines represent those for 
$r_0 = 8$km (Mag-A)  and $5$km (Mag-B), respectively. }
\label{NtStar}
\end{center}
\end{wrapfigure}

We show the baryon density in Fig.~\ref{NtStar}a  and the particle
fractions in Fig.~\ref{NtStar}b  as a function
of  the neutron-star radius. For this figure we assume a neutron star baryonic mass 
of $M_{NS} = 1.68 M_{\odot}$,
a temperature of $T = 20$ MeV, and  a  lepton fraction of $Y_L=0.4$.
The moment of inertia of the neutron star becomes
$I_{NS} = 1.54 \times 10^{45}$ g$\cdot$cm$^2$ or $1.36 \times 10^{45}$
g$\cdot$cm$^2$ with or without Lambda particles, respectively.  We note, however,
that magnetic fields of this strength will also slightly increase the neutron star radius due to the additional magnetic pressure.
The associated increase in the moment of inertia, would therefore decrease  slightly the spin down rate estimated here [see Eq.~(\ref{PdotPN}) below].  
Nevertheless, we ignore this effect as our purpose is to estimate the maximum possible spin-down rate from 
asymmetric neutrino scattering.

One can see in Figs.~\ref{NtStar}a and \ref{NtStar}b  the appearance of Lambdas for a baryon density greater than about twice the saturation density of nuclear matter, i.e. $\rho_B \gtrsim 2 \rho_0$, where $\rho_0 \approx 2.7 \times 10^{14}$ g cm$^{-3}$. 
This softens the EOS and  leads to an 
increase in the baryon density and neutrino fraction for $r \lesssim 8$ km relative to hadronic matter without Lambdas.

%%========================================== 
%%========================================== 
%%========================================== 

\subsection{Toroidal magnetic Field}

The ratio of the total rate of angular momentum loss to  the total power
radiated  by  neutrinos at a given spherical surface $S_N$ is 
\begin{equation}
%\frac{c d L_z}{d x_L} / \frac{d E_T}{d x_L}
\left( \frac{cd L_z/dt}{dE_T/dt} \right)
= \frac{ \int_{S_N} d \Omega_r
\int \frac{d^3 k}{(2 \pi)^3} \Delta f (\vbr,\vk) (\vbr \times \vk )
\cdot {\vn}}
{ \int_{S_N} d \Omega_r
\int \frac{d^3 k}{(2 \pi)^3} \Delta f (\vbr,\vk) \vk \cdot {\vn}} ~~,
\label{AngMomZ}
\end{equation}
where ${\vn}$ is the unit vector normal to $S_N$.  
For illustration we will consider surfaces for which $\rho_B = \rho_0$
and $\rho_B = \rho_0/10$. 
We also adopt the speed of light for the neutrino propagation velocity.
We can then obtain the angular acceleration from the neutrino luminosity,
${\cal L}_{\nu} = (dE_T/dt)$,
\begin{equation}\
{\dot \omega} = - \frac{1}{c I_{NS}}
\left( \frac{cd L_z/dt}{dE_T/dt} \right) {\cal L}_{\nu} ~.
\end{equation}
For a PNS with spin period  $P$, the angular velocity is  $\omega = 2 \pi /P$, 
and the angular acceleration is defined by 
${\dot \omega} = - 2\pi {\dot P}/P^2$.  
Thus, we  obtain
\begin{equation}
\frac{\dot P}{P} = \frac{P}{2 \pi c I_{NS}}
\left( \frac{cd L_z/dt}{dE_T/dt} \right) {\cal L}_{\nu}~~ .
\label{PdotPN}
\end{equation}

We adopt the following parameterization for the toroidal magnetic field 
configuration in cylindrical coordinates $(r_T, \phi,z)$, 
\begin{equation}
\vec{B} = B_{\phi} G_L(z) G_T(r_T) {\hat e}_{\phi},
\end{equation}
where ${\hat e}_{\phi} = (-\sin \phi, \cos \phi, 0)$ in terms of 
the azimuthal angle $\phi$, and
\begin{eqnarray}
G_L(z) = \frac{4e^{z/a_0}}{\left[1 + e^{z/a_0} \right]^2},
&~~~&
G_T(r_T) = \frac{4e^{(r_T-r_0)/a_0}}
{\left[1 + e^{(r_T-r_0)/a_0}  \right]^2} .
\end{eqnarray}
This functional form was chosen to approximate the results of numerical simulations  \cite{KM10,TKS09} of toroidal magnetic field amplification.
For purposes of estimating the maximum possible effect we assume a  
 toroidal magnetic field that is aligned along the direction of 
the spin rotation.  
Admittedly, this is an oversimplification, but it is adequate for our purposes 
of estimating the maximum possible effect.  
Nature, however, could be more complicated.  
Toroidal fields could be oppositely oriented and can even invert
with time.  
This would imply that neutrino emission could also accelerate 
the stellar rotation.
Another plausible case is an (anti-parallel) poloidal torus
configuration of the magnetic field.
In this case more a complicated scenario could be possible.
Assuming that in the northern hemisphere the direction of the magnetic field 
and the spin rotation are the same, 
then the effect described here would would operate to decelerate the rotation.
In the southern hemisphere, however, the magnetic field and spin could 
be antiparallel.  
In this case the asymmetry in neutrino absorption may even accelerate 
the rotation.
This might lead to a complicated twisting mode.

In Fig.~\ref{NtStar}c we illustrate  the magnetic field strength $|{\vB}/B_{\phi}|$ for different field configurations, 
 with $a_0=0.5$ km and  $r_0 = 8.0$ km (Mag-A) or $r_0 = 5.0$ km (Mag-B).  These parameters are chosen to represent 
 a best case and a typical case.  As such, this should bracket the cases for which the effect studied here may be of interest.
 The fact that the spin down is still significant in both limits supports the robustness of these results.
We here take ${\cal L}_\nu \approx 3 \times 10^{52}$ erg$\cdot$s$^{-1}$
\cite{lai98} as a typical value of  the neutrino luminosity from the
proto-neutron star, and the spin period is chosen to be $P=10$ ms, while
the observed spin period of magnetars is about 10 s \cite{PSdata,WT04}.
We discuss below more details regarding this choice of spin period.  

\section{Result and Discussions}

Numerical simulations \cite{KM10,TKS09} have shown that 
the strength of the toroidal magnetic field can easily amplify to 
$B_\phi = 10^{16}$G or more from an initial value of $\sim 10^{14}$G 
due to the winding of the  magnetic field lines in  rapidly rotating of PNSs.  
We therefore adopt these typical values for both components 
$B_{pol} = 10^{14}$G and $B_\phi = 10^{16}$G, respectively.

We summarize the calculated results in Table \ref{Tres1}.  
It includes two cases by taking the PNS surface 
$S_N$ at different locations, one at $\rho_B=\rho_0$ and the other at $\rho_B=\rho_0/10$, to illustrate  the robustness of  this braking mechanism. 
We obtain the results that ${\dot P}/P \sim 10^{-6}$ in Mag-A and  
 ${\dot P}/P \sim 10^{-7}$ in Mag-B when $P=10$ms.  

To compare with the rate of spin down  due \cite{nakano11}  to dissipating rotational energy into magnetic dipole radiation 
the sixth column shows ${\dot P}/P$ calculated with the  magnetic dipole radiation (MDR) 
formula~\cite{MDR}.
\begin{equation}
P {\dot P} = B_{pol}^2 \left( \frac{3 I_{NS}c^3}{8 \pi^2 R^6 }\right)^{-1} 
 = B_{pol}^2 \left( \frac{3 M_{NS}^3 c^3}{125 \pi^2 I_{NS}^2 }\right)^{-1}~~,
 \label{MDRP}
\end{equation}
where $R$ and $I_{NS} = 2M_{NS}R^2/5$ are the NS radius and 
the moment of inertia.  
These quantities are determined from the EOS as discussed above. 
For these particular parameters we see that the spin deceleration
from asymmetric neutrino emission can be more effective than that
of MDR when the neutrino luminosity is high.

\vspace{0.5cm} 
\begin{table}[ht]
\begin{tabular}{|c|c|c|cc|c|c|}
\hline
\multirow{2}{*}{~~Comp.~} & \multirow{2}{*}{~Mag.~}
& \multirow{2}{*}{~$\frac{c d L_z / dt}{dE_T / dt}$~}
& \multicolumn{4}{|c|}{ ~~~ ${\dot P}/P$ ~(s$^{-1}$)~~~ } \\
\cline{4-7}
&&& ~$\rho_B = \rho_0$~~ &  ~~$\rho_B = \rho_0/10$~~ & ~~MDR~~ &~Thompson~ \\
\hline
\multirow{2}{*}{~p, n~} & ~Mag-A~ & ~3.34~ &
~~$3.45 \times 10^{-6}$~~ & ~~$7.25 \times 10^{-7}$~~
&\multirow{2}{*}{~$9.86 \times 10^{-8}$~} 
&\multirow{2}{*}{~$3.56 \times 10^{-3}$~} \\
\cline{2-5} &~Mag-B~ & 0.482 &
~~$4.97 \times 10^{-7}$~~ & ~~$3.16 \times 10^{-7}$~~ & & \\
\hline
\multirow{2}{*}{~p, n, $\Lambda$~} &~Mag-A~& ~5.45~ & 
~~$6.39 \times 10^{-6}$~~ & ~~$1.02 \times 10^{-6}$~~
&\multirow{2}{*}{~$7.76 \times 10^{-8}$~} & 
\multirow{2}{*}{~$3.50 \times 10^{-3}$~}\\
\cline{2-5}
& ~Mag-B~ & ~~0.390~~ & 
~~$4.57 \times 10^{-7}~~$ & ~~$2.01 \times 10^{-7}$~~ & & \\
\hline
\end{tabular}
\caption{\small
The 1st column shows the presumed composition of  nuclear matter, i.e.
 "p,~n" for nucleonic and "p,~n,~$\Lambda$"
for hyperonic matter. 
The 2nd column gives  the model for the toroidal magnetic field
 configuration (see text).  
The 3rd column denotes results from Eq.~(\ref{AngMomZ}),  
the 4th and 5th columns are results obtained using  Eq.~(\ref{PdotPN})
 at the indicated baryon density.  
The 6th column shows the spin-down rate  from magnetic
 dipole radiation, Eq.~(\ref{MDRP}). 
The 7th column shows the spin-down rates from the model of  Thompson et al.~\cite{TCQ04}
for the neutrino-driven winds coupled with the poloidal magnetic field, 
Eq.~(\ref{Thompson}).
The spin period is taken to be $P=10$ ms, 
and  magnetic field strengths of $B_{pol} = 10^{14}$ G
and $B_\phi = 10^{16}$ G are used in these calculations.
}
\label{Tres1}
\end{table}
\vspace{0.5cm}

If we consider the case $P \approx 1$ ms, these two mechanisms give 
comparative results.  
This is because ${\dot P}/P$ is proportional to $P$ in our model, 
while it is proportional to $P^{-2}$ in the MDR according to 
Eq.~(\ref{PdotPN}) and Eq.~(\ref{MDRP}). 
If we consider the alternative case of a stronger poloidal magnetic field 
$B_{pol} = 10^{15}$ G 
while keeping $P = 10$ ms, the two mechanisms also give comparable strength
because ${\dot P}/P$ is proportional to $B$ in our model, 
but to $B^{2}$ in MDR. 
Either  conditions of a longer spin period or a weaker field strength 
would thus lead to a dominance of our new mechanism over MDR.

However, other means to spin down magnetic  neutron stars by neutrino emission have been proposed \cite{TCQ04,Heger05,Ott06}.
Even  non-magnetized  rotating neutron star will lose angular momentum by
neutrino emission 
~\cite{Kazanas77, Epstein78,Baumgarte98,Janka04,Heger05,Ott06}.

For illustration, therefore, we also compare with the spin-down of 
the proto-neutron star as was proposed by Thompson et al.~\cite{TCQ04}. 
This mechanism utilizes the neutrino-driven winds to push magnetically
locked matter away from PNS and slow down the rotation. 
In this mechanism spin-down rate in the dipole magnetic field is 
given by \cite{TCQ04},
\begin{eqnarray}
\left({\dot P}/P\right)_{Poloidal}
&=& 4.14\times10^{-3}[{\rm s}^{-1}] 
\left(\frac{M_{NS}}{1.4 M_{\odot}}\right)^{-1}
\left(\frac{ {\dot M} }{10^{-3} M_{\odot}}\right)^{+3/5}
\nonumber \\ && \quad \times
\left(\frac{R}{10{\rm [km]}}\right)^{+2/5} 
\left(\frac{B_{pol}}{10^{14}{\rm [G]}}\right)^{+4/5}
\left(\frac{P}{10{\rm}{\rm [ms]}}\right)^{+2/5} ,
 \label{Thompson}
\end{eqnarray}
where  ${\dot M}$ is the wind mass loss rate.
A comparison between our rate Eq.~(\ref{PdotPN}) 
from asymmetric neutrino emission and the neutrino driven wind 
Eq.~(\ref{Thompson}) is shown in the seventh column of Table \ref{Tres1}.
For this comparison we use the standard parameter values of 
$M_{NS} = 1.68 M_{\odot}$,  ${\dot M} = 10^{-3} M_{\odot}$, 
$B_{pol} = 10^{14}$G, $P = 10$ ms, 
and $R$ = 10.1 km (with $\Lambda$s) and 10.8 km (without $\Lambda$s),
which are obtained from $I_{NS}$.

One can also estimate the effect from non-magnetic neutrino transport.  
This effect arises \cite{Baumgarte98} when the neutrino transport  
is treated in the comoving frame of the fluid.  
In this corotating frame the neutrino distribution will be isotropic 
in equilibrium in the absence of strong magnetic fields. 
In the laboratory frame, however, the rotation of the neutron
star creates an emission asymmetry by which the neutrinos are able to
carry away angular momentum \cite{Baumgarte98}. 
Based upon Eq.~(15) in Ref.~\cite{Baumgarte98}, one can estimate that  
$\dot P/P \le 2.5 (\dot M/M) < 3 \times 10^{-3}$ s$^{-1}$,  
for $\dot M/M \sim 10^{-3}$ s$^{-1}$ 
as in the wind model above \cite{Baumgarte98}.   
Hence, this mechanism may also exceed or be comparable to the effect 
from asymmetric neutrino scattering described 
in the present work.

Nevertheless, the spin-down mechanism described in the present work 
can be an additional effect which works together 
with the other neutrino-emission mechanisms.
It may cause further enhancement of the spin-down rate, and therefore  
warrants consideration in models for the spin down of the PNS.

We note, however, that it may be difficult to directly confirm by 
observations the asymmetric neutrino scattering mechanism described herein.  
In principle one might eventually confirm this effect via a detection of 
neutrinos aligned or anti-aligned with a magnetic field.  
In this regard, there is another consequence of asymmetric neutrino scattering 
that is more directly observable, i.e.  the observed pulsar kick velocities.
In our previous paper~\cite{MKYCR11,MYCHKMR13}, we showed that 
the neutrino asymmetric emission can lead to  pulsar kick velocities of  
$v_{kick} = 500 \sim 600$ km/s,
that are comparable to the observed values of $400 - 1500$ km/s.
Hence, asymmetric neutrino emission  may affect a variety of observed 
dynamical processes associated with SN explosions.

We note that in n the present calculation we have ignored the neutrino scattering
and production processes.
The neutrino scattering process enhances the asymmetry of the emission, 
although its contribution to the mean-free path is much
smaller than that from absorption in the density region of interest, 
$\rho_0 \lesssim \rho_B \lesssim  3\rho_0$ \cite{MYCHKMR13}. 

Neutrino production in a  magnetic field is known to cause  asymmetry 
in the neutrino emission \cite{chugai84,kisslinger2}.
The cross section for the neutrino production reaction, 
$e^{-} + p \rightarrow n, \Lambda + \nu_e$, 
is qualitatively the same as that for the  absorption reaction,  
$\nu_e + n, \Lambda \rightarrow p + e^{-}$.  
The only difference is  the small contribution from the magnetic part of 
the initial and final electron states. 
Hence, this production process would  tend to enhance  the asymmetry 
and  also contribute to the spin deceleration.

Our goal in this work has been to estimate the maximum possible effect 
of asymmetric neutrino scattering on the spin down rate of PNSs.  
Even so, there are a number of  uncertainties in our estimate of ${\dot P}/P$.  
These include the interior strength and configuration of the magnetic field, 
along with the spin period of the NS core, etc.  
This process may or may not contribute, but should at least be considered 
in a more realistic calculation.
Since our value of ${\dot P}/P$ is at least $10^{2}$ times  larger than 
that for the MDR spin-down mechanism, asymmetric neutrino emission 
could be significant at some point during the early stages of SN explosion.  
Moreover, as discussed above, other processes such as neutrino scattering 
and production tend to increase the asymmetry in neutrino emission and 
lead to additional spin deceleration. 
Thus, we can conclude that  asymmetric neutrino emission from PNSs may play 
a role in the spin deceleration of a magnetic PNS and should be considered.

\section{Summary}

We have estimated  a best case scenario for the possible spin down of a PNS 
due to  asymmetric neutrino absorption.
We consider the optimum case of a toroidal magnetic field
configuration aligned with the neutron star spin direction.  
We calculated  
the cross-sections  for asymmetric neutrino absorption and scattering 
in the context of RMF theory.
We then solved the Boltzmann equation using  a one-dimensional
attenuation method, assuming that the neutrinos propagate along an
approximately  straight line, and  that the system is in quasi-equilibrium.
We  only included  neutrino absorption which dominates \cite{MYCHKMR13}
over  scattering in producing asymmetric momentum transfer to the PNS.

In this simplified model we found that asymmetric neutrino emission 
can have an effect on the early  spin deceleration of a PNS.  
Indeed, this effect can initially be  larger than the braking 
from a magnetic dipole field configuration, but is probably smaller than  
that due to the magnetized neutrino wind breaking mechanism of \cite{TCQ04}. 
Finally, we caution that  definitive conclusions should involve 
a fully dynamical  MHD simulation of the evolution of a PNS with
asymmetric neutrino scattering and production as well as absorption 
in a strong magnetic field. 
Nevertheless, the results presented here suggest  a possible
influence of asymmetric neutrino absorption on the early formation process 
of magnetars and therefore warrant further investigation. 
 
\acknowledgements
%\bigskip
%
This work was supported in part by the Grants-in-Aid for the Scientific
Research from the Ministry of Education, Science and Culture of
Japan~(20105004, 21105512, 2324036, 2354032, 24340060, 25105510)
and Nihon University College of Bioresource Sciences
Research Grant for 2013.
Work (MKC) is supported by the National Research Foundation of Korea
(2012R1A1A3009733, 2011-0015467).
Work at the University of Notre Dame (GJM) is supported
by the U.S. Department of Energy under
Nuclear Theory Grant DE-FG02-95-ER40934.

\end{document}